\documentclass[aps,prb,twocolumn]{revtex4}
\usepackage{epsfig}
\usepackage{amsmath}
\usepackage{subcaption}
\usepackage[bookmarks=false]{hyperref}
\hypersetup{colorlinks=true, citecolor=blue, urlcolor=blue, linkcolor=blue}

\begin{document}

\title{Cooling molecular electronic junctions by AC current }
\author{Riley J. Preston}
\author{Thomas D. Honeychurch}
\author{  Daniel S. Kosov\footnote{E-mail: daniel.kosov@jcu.edu.au}}
\address{College of Science and Engineering,  James Cook University, Townsville, QLD, 4811, Australia }


\begin{abstract}
Electronic current flowing in a molecular electronic junction dissipates significant amounts of energy to  vibrational degrees of freedom,  straining and rupturing chemical bonds and often quickly destroying the  integrity of the molecular device. The infamous mechanical instability of molecular electronic junctions critically limits performance, lifespan, and raises questions as to the technological viability of single-molecule electronics.
Here we propose a practical scheme for cooling the molecular vibrational temperature via application of an AC voltage over a large, static operational DC voltage bias. Using nonequilibrium Green's functions, we computed the viscosity and diffusion coefficient experienced by nuclei surrounded by a nonequilibrium "sea" of periodically driven, current-carrying  electrons. The effective molecular junction temperature is deduced by balancing the viscosity and diffusion coefficients.
Our calculations show the opportunity of achieving in excess of 40\% cooling of the molecular junction temperature while maintaining the same average current.

\end{abstract}

\maketitle
\section{introduction}
The basic building block for molecular electronics is the single-molecule junction: a molecule chemically linked to two leads. To date, the major achievements in molecular electronics have been in the development of a fundamental understanding of the quantum transport in single molecules and ways to control it.  However, there is a key scientific challenge to be overcome before the commercial potential of these technologies can be realized – 
the lifetime of molecular devices is notoriously small \cite{doi:10.1021/ja902871d,doi:10.1021/nl801669e,doi:10.1021/ja074456t}. The record lifetime achieved this year in a breakthrough experiment  is still only 2.7 seconds\cite{D0SC01073A}, which is  obviously much shorter than what is expected for feasible  post-silicon technology.

Electric current flowing from macroscopic leads through a sub-nanometer wide molecular constriction  deposits  significant amounts of power 
into the molecular junction.
Moreover, the situation is exacerbated by the large molecule-metal contact resistance, typically,   1-100 M$\Omega$
which comes from the misalignment of molecular levels and the leads' Fermi energies.
Molecular junctions  behave like insulators and  require a high  voltage bias for device operation.  As a result, the significant operational voltage bias of a few volts across the molecular length along with large electric current densities destroy the molecular device’s structural integrity through chemical bond rupture, large scale molecular geometry alteration or electromigration of the lead interfacial atoms.

The physical mechanisms of current-induced molecular device breakdown have been comprehensively studied  experimentally\cite{doi:10.1021/jacs.6b10700,doi:10.1021/ja512523r,darwish2016,doi:10.1021/acs.nanolett.9b00316}, and theoretically\cite{peskin2018,peskin18,fuse,dundas12,C2CP40851A,dzhioev11,thoss11,catalysis12,preston2020,pistolesi10}.
Recent theoretical proposals to specially engineer energy dependence of the lead density of states \cite{peskin18} seem to lack practical appeal. The idea of using spin-polarised current to cool  vibrations \cite{PhysRevLett.113.076602,PhysRevLett.113.047201}  is limited to the use of ferromagnetic leads which makes it inapplicable to the vast majority of molecular electronic devices. Peskin et al. have recently proposed to use electrode plasmon excitations in electrodes to reduce power dissipation in molecular junctions \cite{peskin2020}. {   The possibility to reduce the heating by increasing the ambient  temperature of the device has been recently demonstrated theoretically \cite{PhysRevB.98.081404}.  }
Despite all of these efforts, a practical solution to the sensitivity of structural stability in molecular junctions remains elusive. Subsequently, we propose a new strategy to decrease the Joule heating in molecular junctions: the application of a sinusoidal voltage over the large DC voltage bias which acts to reduce the effective vibrational temperature of the molecular junction.

We set $\hbar=1$ in our derivations.

\section{Model hamiltonian}
The molecular junction is described by the general tunnelling Hamiltonian
\begin{equation}
H = H_M + H_{ML} + H_{MR} + H_{L} + H_{R},
\end{equation}
where $H_M$ is the Hamiltonian for the molecule, $H_L$  and  $H_R$ are the Hamiltonians for the left and right leads, while  $H_{ML}$  and $H_{MR}$ are for the interaction between the central region and the left and right leads, respectively. 

The molecule is modelled as a molecular orbital  coupled to  a single classical  degree of freedom
\begin{equation}
H_M= \epsilon(x) d^\dag d +\frac{p^2}{2m} +U(x).
\end{equation}
Here $\epsilon(x) $ is the energy of the molecular orbital. It is a function of the classical coordinate $x$,  which along with the corresponding momentum $p$ and potential $U(x)$ describe the molecular geometry. 
Operator $d^\dag$ ($d$) creates (annihilates) an electron in the molecule. 
We assume in our calculations that the molecular orbital depends linearly on $x$
\begin{equation}
\epsilon(x) = \epsilon_0 + \lambda x,
\label{ex}
\end{equation}
where $\lambda$ is the coupling strength between the electronic and nuclear degrees of freedom.
The classical potential is taken in the harmonic oscillator form
\begin{equation}
U(x) =\frac{1}{2} k x^2,
\end{equation}
where $k$ is the spring constant associated with the chemical bond of interest. Notice that $x=x(t)$ is a stochastic, time-dependent variable which satisfies the Langevin equation with noise and viscosity obtained from nonequilibrium Green's functions (NEGF) calculations described below.
The spin of electrons does not play any physical role here and will not be included explicitly into the equations.

Both  leads  are modelled as macroscopic reservoirs of noninteracting electrons
\begin{equation}
\label{leads}
H_L +H_R =  \sum_{ k\alpha=L,R}  \epsilon_{k\alpha}(t) a^{\dagger}_{k\alpha} a_{k\alpha},
\end{equation}
where $a^{\dagger}_{k\alpha}$($a_{k\alpha}$) creates (annihilates) an electron in the single-particle state $k$ of either the left ($\alpha=L$) or  the right ($\alpha=R$) lead. 
The leads energy levels have a sinusoidal dependence on time due to an external AC driving with
frequency $\Omega$ and amplitude $\Delta_\alpha$
\begin{equation}
	\epsilon_{k\alpha} (t) = \epsilon_{k\alpha} + \Delta_\alpha \cos (\Omega t).
\end{equation}
Additionally, the leads are also held at different static chemical potentials $\mu_\alpha$ at  all times, the difference between them corresponds to the applied DC voltage bias 
$V=\mu_L-\mu_R$.
Both sinusoidal AC and static DC voltages are applied symmetrically in our calculations:  $\Delta_L= -\Delta_R$ and $\mu_L=-\mu_R$, 
and the leads' Fermi energies are set to zero.

The coupling between the central region and the left and right leads   are  given by the tunnelling interaction
\begin{equation}
\label{coupling}
H_{ML}+H_{MR}=  \sum_{k \alpha=L,R } (t_{k\alpha} a^{\dagger}_{k \alpha} d +\mbox{h.c.} ),
\end{equation}
where $t_{k\alpha}$ is the tunnelling amplitude between the leads  single-particle states and the molecular orbital.

\section{Nonequilibrium viscosity and noise  produced by AC driven electrons}

Joule heating  is a balancing process between the electronic time-dependent viscosity and the amplitude of the  
random force  exerted by AC driven electrons. They play opposite roles:  the viscosity deposits energy from nuclear vibrations back onto the electrons, while the random force dissipates the power of the electric current into nuclear motion. If one prevails, it results in the domination of cooling or heating processes.

The time-dependent electronic viscosity and diffusion coefficients produced by the AC driven electrons are computed using NEGF.
This method is based on the  time-separation solution of the Keldysh-Kadanoff-Baym equations for nonequilibrium Green's functions,  and utilizing this to compute the force exerted by the electrons on the nuclear degrees of freedom and the time correlation of the dispersion of the force operator \cite{Bode12,preston2020,subotnik17-prl,subotnik17,subotnik18}.
 These derivations follow directly the derivations of the viscosity and diffusion for a system with time-dependent coupling to the leads as discussed in \cite{preston2020,doi:10.1063/1.4965823}, as long as the time-derivatives of the self-energies are not specified explicitly.
Notice that in addition to the standard assumption that the dynamics of mechanical degrees of freedom are slow in comparison with the electron tunneling time, we have to assume that
 the rate of change of the leads' energies  due to the external AC driving is  also smaller than the electronic tunneling time across the junction \cite{honeychurch2020}. This means 
that the validity of our approach requires that 
the characteristic frequency of the nuclear motion and the AC driving frequency should both be smaller than the molecular level broadening due to the coupling to the leads.

\begin{widetext}
Under these separable time-scale assumptions, the viscosity $\xi$ and the diffusion coefficient $D$ can be computed from the Keldysh-Kadanoff-Baym equation and are given by the following expressions (details of the derivations are shown in supplementary materials):
\begin{equation}
\xi(t) = \frac{[\epsilon'(t)]^2}{2} \; \int   \frac{d\omega}{2 \pi}   {G}^{R}(t,\omega)     {G}^{A}(t,\omega) \Big(    {G}^{A}(t,\omega)  -    {G}^{R}(t,\omega) \Big) \partial_{\omega} \Sigma^{<} (t,\omega),
\label{xi}
\end{equation}
\begin{equation}
D(t) =  [\epsilon'(t)]^2  \int     \frac{d\omega}{2 \pi}    
 G^<(t,\omega) G^>(t,\omega) + [\epsilon'(t)]^2  \int     \frac{d\omega}{2 \pi}    
\Big\{ \delta G^<(t,\omega) G^>(t,\omega) +    
 G^<(t,\omega)\delta G^>(t,\omega) \Big\}.
 \label{D}
\end{equation}
\end{widetext}

The viscosity and diffusion coefficients depend on the advanced and retarded adiabatic Green's functions,
\begin{equation}
    {G}^{A/R}(t,\omega)
=
\left[ \omega - \epsilon (t) - \Sigma^{A/R}(t,\omega)  \right]^{-1},
\end{equation}
\begin{equation}
    {G}^{</>}(t,\omega)
=
   {G}^{R}(t,\omega) \Sigma^{</>}(t,\omega) G^{A}(t,\omega) 
\end{equation}
which follow the
 adiabatically, instantaneously computed time-dependent trajectories of the mechanical degrees of freedom as well as the external AC driving of the leads. The diffusion coefficient  also depends on the
 first order dynamical corrections to the lesser and greater Green's functions  due to the AC driving in the leads' self-energies
 \begin{multline}
 \delta{ {G}}^{</>}(t,\omega) = \frac{i}{2}  {G}^{R}(t,\omega)   {G}^{A}(t,\omega) 
 \\
 \times
 \Big(  {G}^{A}(t,\omega)  -  {G}^{R}(t,\omega) \Big)
\partial_{t} \Sigma^{</>}(t,\omega).
 \end{multline}
  
In all our calculations we employ the wide-band approximation for the leads' density of states and tunnelling amplitudes,
which results in the following expressions for the self-energies of the 
AC driven leads:
	\begin{multline}
	 \Sigma^<_{\alpha}(t,\omega)=
	i \Gamma_\alpha \sum_{n=-\infty}^{\infty} f_\alpha(\omega+ n\Omega/2) \\
	\times
	(-1)^n J_n\left(\frac{2\Delta_\alpha \cos(\Omega t)}{\Omega}\right)
	\end{multline}
	\begin{equation}
	 \Sigma^>_{\alpha}(t,\omega)
	= 
	-i \Gamma_\alpha + \Sigma^<_{\alpha}(t,\omega),
	\end{equation}
	\begin{equation}
	 \Sigma^A_{\alpha}=
	\frac{i}{2} \Gamma_\alpha; \;\;\; \Sigma^R_{\alpha}=
	-\frac{i}{2} \Gamma_\alpha,
	\end{equation}
where $\Gamma_\alpha$ is the standard level-broadening function due to the coupling to lead $\alpha$, $f_\alpha(\omega)$ is the Fermi-Dirac distribution of electronic occupations in the $\alpha$ lead, and  
 $J_n(z)$ is the Bessel function of the first type.
 {  
 The total level broadening
 \begin{equation}
 \Gamma=\Gamma_L + \Gamma_R
 \end{equation}
 will be used  as an energy scale to represent all model parameters used in the calculations. }

The expression for the viscosity (\ref{xi}) resembles the viscosity for a system with DC current \cite{Bode12,preston2020,subotnik17-prl,subotnik17,subotnik18}, however, it  is defined  here using Green's functions with sinusoidally  driven lead self-energies. The diffusion coefficient (\ref{D})
consists of two terms: the first term is again the standard expression similar to what was used in DC current junctions \cite{Bode12,preston2020,subotnik17-prl,subotnik17,subotnik18}, the second term is new and arises from the dynamical corrections to the lesser and greater Green's functions (again computed using sinusoidally oscillating self-energies).

\begin{figure}
\centering
\begin{subfigure}{0.47\textwidth}
\centering
\includegraphics[width=1\textwidth]{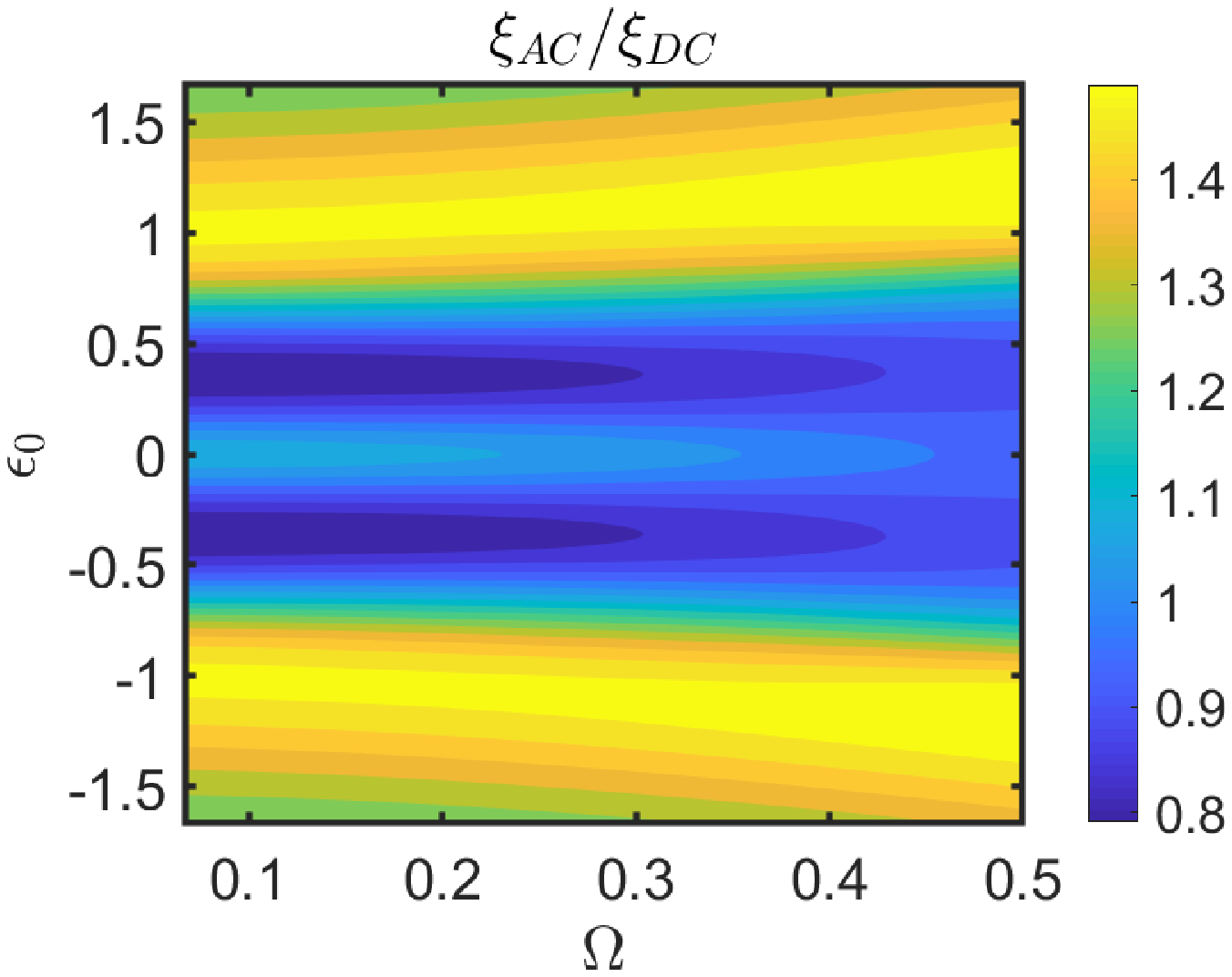}
\caption{}
\label{1a}
\end{subfigure}
\begin{subfigure}{0.47\textwidth}
\centering
\includegraphics[width=1\textwidth]{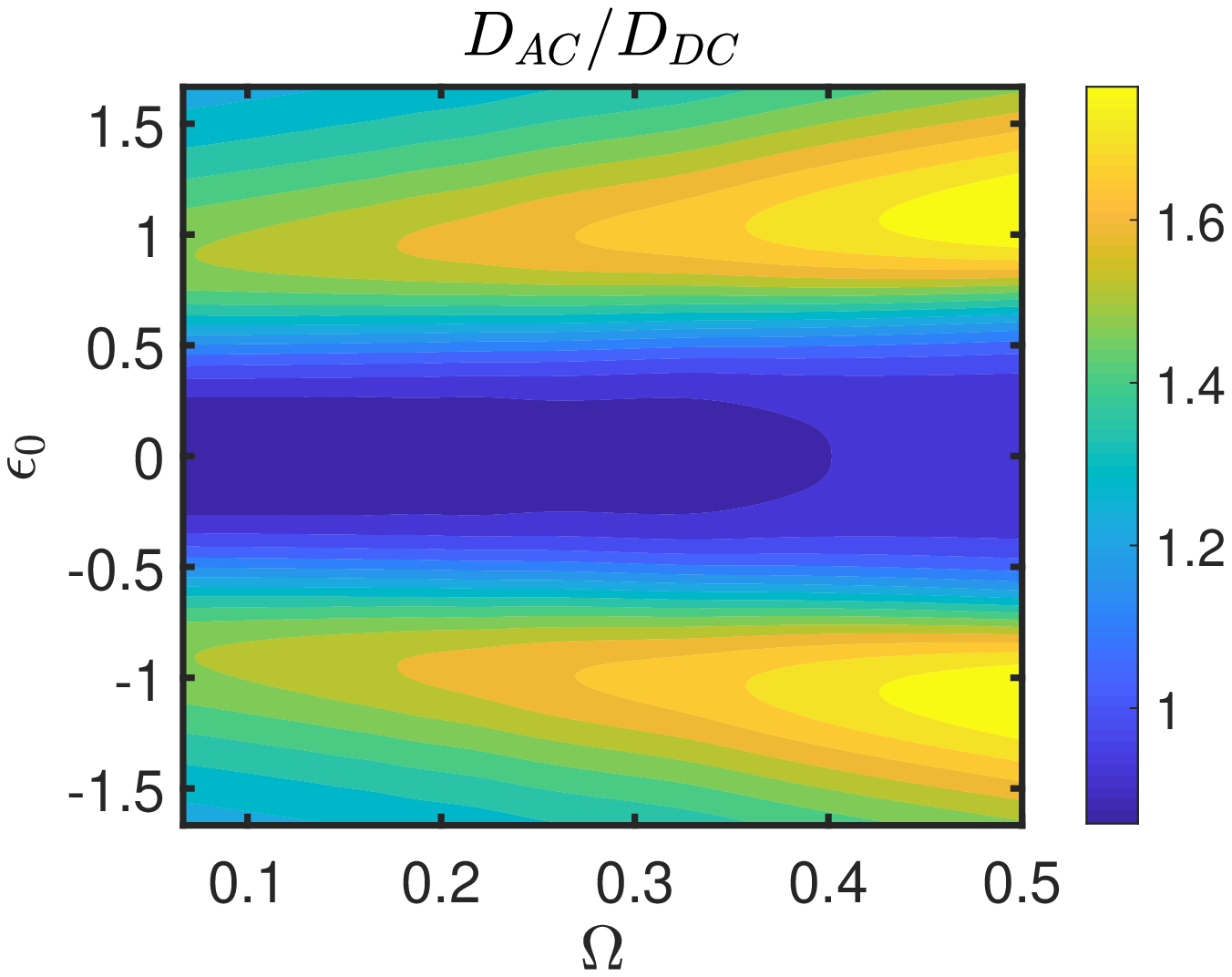}
\caption{}
\label{1b}
\end{subfigure}

\caption{Electronic viscosity (a) and diffusion coefficient (b) computed as functions of the AC driving frequency and the molecular orbital energy.  Parameters used in calculations: $\lambda^2 / k = 0.002 \Gamma $, $V=2 \Gamma /3$, $\Delta=\Gamma/3$.  $\Omega$ and $\epsilon_0$ are given in terms of $\Gamma$.}
\label{fig1}
\end{figure}

 Figure \ref{fig1} shows the viscosity and diffusion coefficient (averaged over a period of oscillation and also statistically averaged with respect to possible values of $x$). The viscosity and diffusion coefficients are shown as a ratio to the corresponding DC (static) values for a given average voltage, the DC  calculations are performed using   (\ref{xi}) and (\ref{D})  and setting the amplitude of sinusoidal voltage modulation $\Delta=0$  \cite{preston2020}. 
{    The leads temperature is  set to  $0.02\Gamma$ in all our calculations.}
In the centre of the resonance region, application of the AC driving acts to decrease the diffusion coefficient while slightly increasing the viscosity. In our previous work \cite{preston2020}  (and in agreement with Subotnik et al results \cite{doi:10.1063/1.5035412}), peaks in the friction occur when the molecular orbital energy aligns with either the left or right chemical potentials, since the electrons can deposit any amount of energy taken from the nuclear degrees of freedom to the leads via inelastic scattering to the available empty states above the chemical potential. 
This is in contrast to the diffusion which  has contributions from all lead states in the resonant region. Applying this analysis to our system, we observe that the viscosity increase in the resonant region is a result of the lead chemical potentials being allowed to shift closer to the resonant level and inducing increased interaction between the nucleus and the high-energy electrons in the leads. However, the AC voltage  has minimal effect on increasing the diffusion near the resonance. The growth of the viscosity relative to the diffusion coefficient results in an optimal transport regime in which the molecular junction is cooled relative to the static case.
As we shift our resonant level to the edges of the resonant region, we observe a notable decrease to the viscosity upon application of the AC leads, relative to the DC case. In the static regime, the viscosity is maximised here due to the alignment between the resonant level and the leads, which is broken upon application of an AC voltage. Resultantly, the AC driving acts to increase the junction temperature in this region.

\begin{figure}
\includegraphics[width=0.47\textwidth]{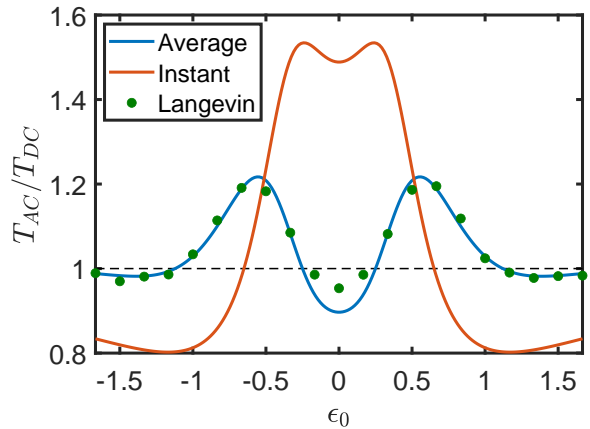}
\caption{Comparison of methods of effective temperature calculation with Langevin-simulated nuclear temperature results for the same parameters.
 Parameters used in calculations: $\lambda^2 / k=\Gamma/6$, $\Omega= \Gamma/15$, $V=2 \Gamma/3$ and $\Delta=\Gamma/3$.  $\epsilon_0$ is given in terms of $\Gamma$.}
\label{newfig}
\end{figure}

{  
\section{Effective temperature }

Using viscosity $\xi$ and diffusion coefficient $D$ we define an effective temperature of the molecular junction via analogy with the equilibrium fluctuation-dissipation theorem as given by $\frac{D}{2\xi}$ for one-dimensional motion. There are two options in defining the effective temperature for a given frequency; we can calculate the instantaneous effective temperature for each given time in a period of AC leads oscillation and then average over the period, given by

\begin{equation}
T^{inst}_{AC}= \frac{\Omega}{2 \pi} \int_0^{2 \pi /\Omega} dt   \frac{D(t)}{2\xi(t)},
\end{equation}
or alternately, we can first take $D$ and $\xi$ to be time-averaged quantities over a period of AC oscillation, then calculate the effective temperature as 

\begin{equation}
T^{ave}_{AC}=  \frac{\int_0^{2 \pi /\Omega} dt D(t)}{2\int_0^{2 \pi /\Omega} dt \xi(t)}.
\end{equation}

In Figure \ref{newfig}, we compare these options with average temperature  data  obtained from kinetic energy calculated via numerical Langevin simulations for the same parameters. In the interest of computational efficiency, $\lambda^2 /k = \Gamma/6$, where all other parameters coincide with other results presented in this study. We observe $T_{AC}^{ave}$ to be a far more accurate measure of the average nuclear temperature within the system for these parameters. Given that $\xi$ and $D$ are each proportional to $\lambda^2$, decreasing $\lambda^2 / k$ (in line with the presented results in this paper) will only further improve the accuracy of the average calculation, since the molecule will react more slowly to temperature variations due to the AC leads oscillations. As such, we choose to use $T^{ave}_{AC}$ as our measure for the effective temperature.

}

\begin{figure}
\centering
\begin{subfigure}{0.47\textwidth}
\centering
\includegraphics[width=1\textwidth]{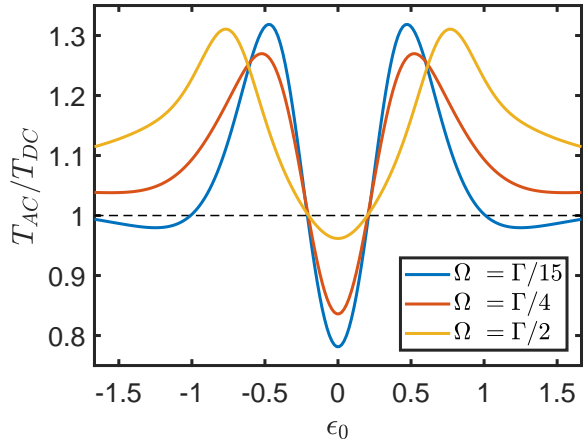}
\caption{}
\label{2a}
\end{subfigure}
\begin{subfigure}{0.47\textwidth}
\centering
\includegraphics[width=1\textwidth]{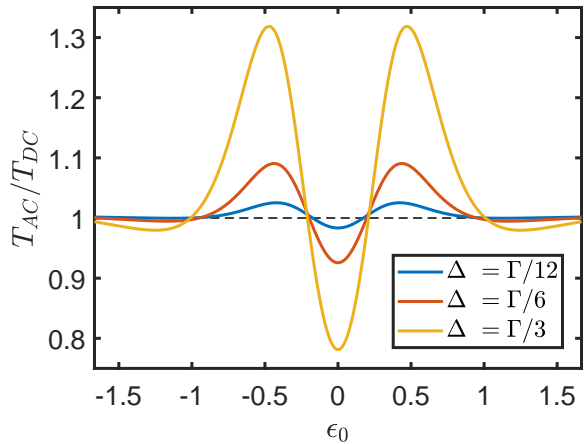}
\caption{}
\label{2b}
\end{subfigure}
\begin{subfigure}{0.47\textwidth}
\centering
\includegraphics[width=1\textwidth]{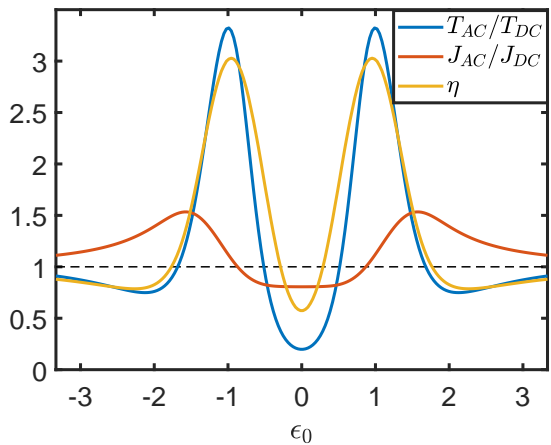}
\caption{}
\label{2c}
\end{subfigure}

\caption{Ratio of AC and DC molecular temperatures computed as functions of molecular orbital energy.
(a)  shows the results for different AC driving frequencies with $\Delta=\Gamma/3$ and $V=2 \Gamma/3$
(b)  shows the results for different  amplitudes of AC voltage oscillations with  $\Omega= \Gamma/15$ and $V=2 \Gamma/3$. 
(c)  shows the temperatures ratio, currents ratio, and cooling ratio defined by Eq.(\ref{eta})   for a higher DC voltage $V=5 \Gamma/3$ and $\Delta=5\Gamma /6$.
 $\lambda^2 / k=0.002 \Gamma$ and $\epsilon_0$ is given in terms of $\Gamma$.}

\label{fig2}
\end{figure}

Figure \ref{fig2} shows the ratio $T_{AC}/T_{DC}$ computed for various transport regimes.
The AC temperature is compared to the static DC temperature  $T_{DC}$ computed again as the ratio between the diffusion and friction coefficients, but now obtained using NEGF calculations for static leads \cite{preston2020}.  Both temperatures are again statistically averaged over $x$ and over a period of AC driving.  As we have already deduced from the behaviour of the viscosity and diffusion coefficient, cooling is observed in the central resonance transport regime, while heating is observed at the edges of resonant transport. The effect of cooling is more significant for slow AC driving  (Figure \ref{fig2}-(a)) and is amplified if the amplitude of voltage driving is increased (Figure \ref{fig2}-(b)). Figure \ref{fig2}-c shows the case of a large static bias voltage, which also enables the consideration of large AC voltage amplitudes; as one observes, it enables a reduction to the effective temperature by as much as 80\% for the chosen parameters, {   while the corresponding decrease to the average current is less than 20\% for the same $\epsilon_0$.}

\begin{figure}
\includegraphics[width=0.47\textwidth]{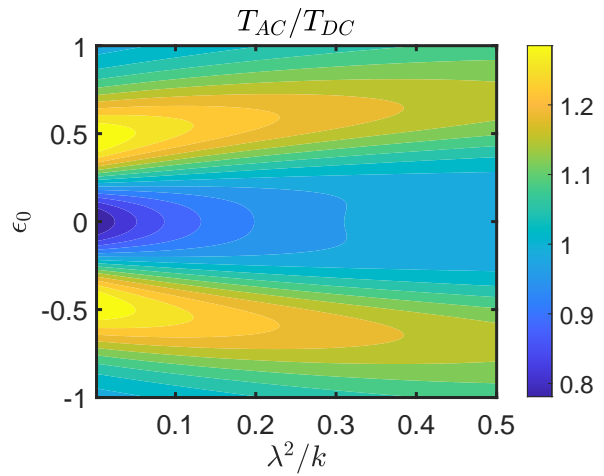}
\caption{Ratio of AC and DC molecular temperatures as function of  $\lambda^2/k$   and $\epsilon_0$. Parameters used in calculations: $\Omega= \Gamma/15$, $V=2 \Gamma/3$ and $\Delta=\Gamma/3$.  Both $\lambda^2/k$   and $\epsilon_0$ are given in terms of $\Gamma$.}
\label{fig3}
\end{figure}

{   
Figure \ref{fig3} demonstrates the role of the chemical bond spring constant $k$ and the coupling strength between the electronic  population and the nuclear motion.  These parameters are interconnected. The term $\lambda x$ in (\ref{ex}) results in the shift of the molecular orbital energy due to deviations away from the equilibrium nuclear position. $\lambda$ describes the magnitude of this shift whilst $k$ governs the range of variation in the $x$ coordinate. Therefore,  $\lambda^2/k$ is an energy  related quantity which encapsulates both effects.
As shown in figure \ref{fig3}, the cooling effects are observed in the resonance regime when $\lambda^2/k <0.2 \Gamma$. This means that this cooling phenomenon can be observed for systems with rigid chemical bonds, or weak electron-nuclear coupling. In any other case, the deviations in the energy level due to nuclear motion may be large enough such that the level leaves this cooling region.
}

The temperature change {\it per se} for a given average voltage may not be a complete measure of heating/cooling, since the AC driving may simply produce a smaller current (averaged over the period of oscillation) relative to the corresponding DC voltage, resulting in less power dissipated over the molecule. 
It is illuminating for us to then consider the heating/cooling effects upon application of an AC driving, relative to the DC case at a given average current (but now a different average voltage).
As such, the static lead electric current $J_{DC}$ is computed using the Landauer formula for static leads, and $J_{AC}$ is the exact electric current (averaged over a period of oscillation) computed using Jauho, Meir, and Wingreen NEGF theory for AC driven quantum transport\cite{Jauho94}. Then we introduce the following quantity called the "cooling ratio"
\begin{equation}
\eta = \frac{T_{AC}(J)}{T_{DC}(J)},
\label{eta}
\end{equation}
which provides a measure of the heating/cooling observed upon application of an AC driving, for a given average current; $\eta < 1$ means that the AC driving yields a lower temperature while allowing for the same average current.
Figure \ref{2c} shows that the application of an AC driving allows for in excess of $40\%$ cooling of the molecular junction, while maintaining the same average current as in the DC case.

\section{Conclusions}

We have demonstrated that the application of an AC driving in the leads' voltage can result in a significant reduction to the power dissipation in a molecular junction, relative to the case of a large static voltage.
The lifetime of a chemical bond is $\tau_{\text{life}} \sim e^{E_b/kT}$, where $E_b$ is the energetic barrier for bond dissociation. One observes that the lifetime depends exponentially on  the effective temperature  $T$; therefore, even a moderate  temperature reduction produces a colossal extension of the device lifetime.
 {    The observed effect is quite robust and does not require special fine-tailoring of the model parameters. Moreover, 
using a master equation derived in the time-averaged Born-Markov approximation  and  assuming  that the driving period must be  shorter than characteristic electron tunneling time $1/\Gamma$,  Peskin  et al. demonstrated that the harmonically driven leads may reduce the vibrational temperature of the molecular junction \cite{peskin2020} . The approach of Ref.\cite{peskin2020} is complementary  in all respects to what we consider in this paper regarding transport and AC driving regimes.  This  serves as a strong indication that the proposed effect is very robust, ubiquitous, and may be applicable for various transport scenarios.}
Although the cooling was the main focus of our paper, it has not escaped our notice that depending upon the parameters, the sinusoidal driving of the leads may result in significant heating of the molecular junction. However, this may also allow for enhanced device functionality as this parameter-controlled heating may be utilized for current-induced
selective bond breaking, and energy efficient single-molecule catalysis of chemical reactions.
\\
\\
\\
\\
\begin{center}
{\bf SUPPLEMENTARY MATERIAL}
\end{center}

Supplementary material describes  detailed derivations of main equations presented in the paper.
\\
\\
\\
\\
\begin{center}
{\bf DATA AVAILABILITY}
\end{center}

The data that supports the findings of this study are available within the article.

\clearpage


\end{document}